\documentclass[aps, preprint, superscriptaddress]{revtex4-2}
\usepackage{graphicx}
\usepackage{makecell, multirow}
\usepackage{color, bm, soul, xcolor}
\usepackage{amsmath, amssymb}
\usepackage{gensymb}
\usepackage{geometry}
\usepackage[colorlinks=true, linkcolor=blue, citecolor=blue, urlcolor=blue]{hyperref}
\newcommand{\etal}{{\em et\,al.}}
\newcommand{\Ed}{{$\mathcal{E}_{\rm d}$}}

\begin{document}

\title{Origin of Anomalous Size Effects in Ferroelectric Hafnia Thin Films}

\author{Tianyuan Zhu}
\email{zhutianyuan@westlake.edu.cn}
\affiliation{Department of Physics, School of Science, Westlake University, Hangzhou, Zhejiang 310030, China}
\affiliation{Institute of Natural Sciences, Westlake Institute for Advanced Study, Hangzhou, Zhejiang 310024, China}

\author{Jingxuan Li}
\affiliation{State Key Laboratory of Precision Welding and Joining of Materials and Structures, School of Materials Science and Engineering, Harbin Institute of Technology, Shenzhen, 518055, China}

\author{Zuhuang Chen}
\affiliation{State Key Laboratory of Precision Welding and Joining of Materials and Structures, School of Materials Science and Engineering, Harbin Institute of Technology, Shenzhen, 518055, China}

\author{Shi Liu}
\email{liushi@westlake.edu.cn}
\affiliation{Department of Physics, School of Science, Westlake University, Hangzhou, Zhejiang 310030, China}
\affiliation{Institute of Natural Sciences, Westlake Institute for Advanced Study, Hangzhou, Zhejiang 310024, China}

\begin{abstract}
{The persistence of ferroelectricity in ultrathin HfO$_2$ films challenges conventional theories, particularly given the paradoxical observation that the out-of-plane lattice spacing increases as the film thickness decreases, an anomalous size effect absent in perovskite ferroelectrics. Here, we resolve this puzzle by revealing that this lattice expansion is counterintuitively coupled to a suppressed out-of-plane polarization. First-principles calculations combined with analytical modeling identify two mechanisms behind this expansion: a negative longitudinal piezoelectric response to the residual depolarization field and a positive surface stress that becomes significant at reduced thickness. Their interplay quantitatively reproduces the experimentally observed lattice expansion. Furthermore, (111)-oriented HfO$_2$ films can support out-of-plane polarization even under open-circuit conditions, in contrast to (001) films that stabilize a nonpolar ground state. This behavior points to the emergence of orientation-induced hyperferroelectricity, an unrecognized mechanism that enables polarization persistence through orientation engineering without electrode screening. We further demonstrate that this principle generalizes to conventional perovskites such as PbTiO$_3$, offering a strategy to eliminate the critical thickness limit by choosing the appropriate film orientation. As a practical pathway to device integration, we also identify the two-dimensional electride Ca$_2$N as a near-ideal electrode that fully restores the ferroelectric properties of HfO$_2$ in ultrathin capacitors.}
\end{abstract}

\maketitle

\clearpage

Ferroelectric hafnia (HfO$_2$) has emerged as a promising candidate for high-density ferroelectric memories, owing to its stable polarization down to one-nanometer scale~\cite{Cheema20p478,Cheema22p648} and excellent compatibility with silicon technology~\cite{Kim21peabe1341,Schroeder22p653}. In conventional perovskite-based ferroelectrics, the depolarization field in a thin film increases with decreasing thickness under the imperfect screening hypothesis, resulting in a critical thickness below which polarization disappears~\cite{Junquera03p506}. In contrast, HfO$_2$-based ferroelectric thin films are largely resistant to this depolarization effect. Ultrathin HfO$_2$ films not only circumvent a ferroelectric critical thickness, but also exhibit enhanced out-of-plane lattice expansion as thickness decreases~\cite{Cheema20p478}; both behaviors constitute anomalous size effects absent in perovskite ferroelectrics.

A direct manifestation of this size effect is the increase in the $d_{111}$ spacing with decreasing thickness in (111)-oriented hafnia thin films~\cite{Cheema20p478}. This structural evolution has often been interpreted as evidence of enhanced ferroelectric polarization at the nanoscale, attributed to a phase transition from the paraelectric tetragonal ($T$, $P4_2/nmc$) phase to the ferroelectric orthorhombic ($O$, $Pca2_1$) phase~\cite{Cheema20p478,Cao24p1080}. However, this interpretation is puzzling, as the ferroelectric $O$ and paraelectric $T$ phases of HfO$_2$ exhibit nearly identical lattice parameters, which typically results in almost completely overlapping XRD reflections for the $O$(111) and $T$(111) planes~\cite{Shiraishi16p262904}. An alternative theoretical explanation attributes the observed lattice expansion to large in-plane biaxial strains, approximately 3.2\% tensile or 5.8\% compressive~\cite{Cao24p1080}. However, the requisite large strains lack experimental evidence, particularly in films grown via typical domain-matching epitaxy~\cite{Estandia20p3801,Fina21p1530}. Specifically, for films grown on highly mismatched substrates, where $m$ film lattices coincide with $n$ substrate lattices, the resulting $m$/$n$ domain-matching arrangements effectively reduces the lattice mismatch and alleviates epitaxial strain~\cite{Fina21p1530}. As a result, the in-plane lattice is only weakly constrained, resembling a relaxed rather than a fully clamped condition. These considerations suggest that the origin of the anomalous out-of-plane lattice expansion in ultrathin hafnia films remains an unresolved question.

The origin of this robust ferroelectricity in ultrathin hafnia films has been explored through several theoretical frameworks, but a complete understanding remains lacking. For instance, the flat phonon band theory proposes that intrinsically localized and individually switchable dipoles account for the persistence of ferroelectricity at the subnanometer scale~\cite{Lee20p1343}. This theory mainly explains the weak coupling of local dipoles in directions perpendicular to the polarization, rather than directly addressing the film's resilience to the out-of-plane depolarization field. Zhou \etal~suggested that coupling between a polar mode and an antipolar mode, which is insensitive to the depolarization field, could stabilize ferroelectricity~\cite{Zhou22peadd5953}. Yet, this mechanism does not explicitly predict the absence of a critical thickness or explain the out-of-plane lattice expansion. Furthermore, these theoretical studies have largely focused on unit cells with polarization axis along the [001] direction, whereas experimentally realized films are most often grown in the (111) orientation~\cite{Cheema20p478,Wei18p1095,Yun22p903}, limiting the direct applicability of these insights.

In this study, we integrate density functional theory (DFT) calculations with Landau--Ginzburg--Devonshire (LGD) theory and surface elastic model analysis to elucidate the origin of the experimentally observed anomalous size effects. We find that HfO$_2$(111) films can evade the conventional critical thickness limit even under open-circuit boundary conditions. However, their polarization is still reduced by the unscreened depolarization field. Counterintuitively, this suppression is accompanied by an out-of-plane lattice expansion, which we attribute to two key mechanisms: an intrinsic negative longitudinal piezoelectric response and a thickness-dependent surface stress effect. Together, these effects quantitatively reproduce the experimentally observed thickness dependence of the out-of-plane lattice spacing. Using the LGD model with DFT-derived material parameters, we show that the persistent out-of-plane polarization in HfO$_2$(111) films is a manifestation of orientation-induced hyperferroelectricity. We further demonstrate that this principle generalizes to conventional perovskites such as PbTiO$_3$, which can sustain out-of-plane polarization even in the absence of electrode screening. This establishes orientation engineering as a general strategy for eliminating the critical thickness limit in conventional perovskite ferroelectrics.

\begin{figure}[t]
\includegraphics[width=6.72 in]{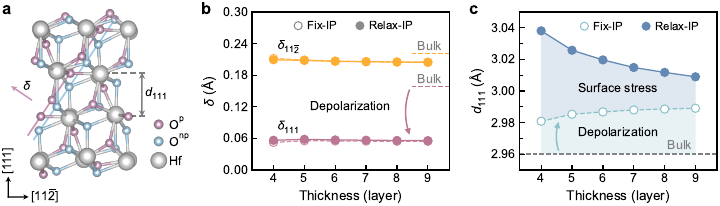}
\caption{\textbf{Absence of critical thickness in ferroelectric HfO$\boldsymbol{_2}$(111) films under open-circuit boundary conditions.} (\textbf{a}) Atomic structure of a fully relaxed four-layer polar $Pca2_1$ (111) slab viewed along [1$\bar{1}$0]. Light blue and purple lines mark the nonpolar oxygen (O$^{\rm np}$) and polar oxygen (O$^{\rm p}$) arrays, respectively. The polar displacement $\delta$ of an oxygen atom is defined relative to the center of its surrounding Hf$_4$ tetrahedron. The out-of-plane lattice spacing $d_{111}$ is determined by the distance between adjacent Hf layers. (\textbf{b}) Polar displacements in slabs of varying thicknesses, with in-plane lattice parameters constrained to the bulk value (fix-IP) or allowed to relax (relax-IP). The in-plane component $\delta_{11\bar{2}}$ remains near the bulk value, whereas the out-of-plane component $\delta_{111}$ is reduced to about one-third of its bulk value by the unscreened depolarization field. (\textbf{c}) Thickness dependence of the out-of-plane lattice spacing $d_{111}$ under fix-IP and relax-IP conditions. Depolarization induces lattice expansion along [111], which is further enhanced by surface stress.}
\label{fig1}
\end{figure}

We start by investigating the size dependence of ferroelectricity in HfO$_2$(111) films under the extreme scenario of an open-circuit boundary condition (OCBC). As the fluorite structure consists of stoichiometric layers stacked along the [111] crystallographic direction~\cite{Tasker79p4977}, we construct HfO$_2$(111) slabs with varying numbers of such layers to systematically explore size effects. Figure~\ref{fig1}a shows the fully relaxed atomic structure of a four-layer polar orthorhombic $Pca2_1$ slab, featuring alternating nonpolar and polar oxygen arrays~\cite{Ma23p256801}. For an oxygen atom between two adjacent Hf layers, the local polar displacement $\delta$ is defined relative to the center of its surrounding Hf$_4$ tetrahedron. The ``bulk region" is designated as the section bounded by the top and bottom subsurface Hf layers. Under this definition, the four-layer slab of 0.9 nm represents the thinnest configuration with a well-defined bulk-like interior. The net polar displacement of a slab is computed by averaging $\delta$ over all oxygen atoms within the bulk region, excluding surface layers to minimize surface-related artifacts, while the interplanar lattice spacing $d_{111}$ is defined as the vertical separation between neighboring Hf atomic planes along the [111] direction.

The unscreened depolarization field directly suppresses polar displacements. As plotted in Fig.~\ref{fig1}b, the out-of-plane polar displacement, $\delta_{111}$, reduces from its bulk value of 0.16~\AA~to just 0.06~\AA. In contrast, the in-plane component, $\delta_{11\bar{2}}$, remains largely unaffected and close to its bulk value of 0.22~\AA. This anisotropic response indicates that, although $\delta_{111}$ and $\delta_{11\bar{2}}$ are geometrically coupled through the (111) orientation, they exhibit markedly different sensitivities to the depolarization field. Importantly, the polar displacement values remain nearly identical in the fix-IP and relax-IP configurations, where fix-IP corresponds to calculations with the in-plane lattice constants constrained to their bulk values, while relax-IP allows full relaxation of both atomic positions and in-plane lattice parameters. This insensitivity indicates that the suppression of $\delta_{111}$ is primarily driven by the depolarization field rather than in-plane strain.

Counterintuitively, this suppression of out-of-plane polarization is accompanied by an anomalous lattice expansion along the same direction. As shown in Fig.~\ref{fig1}c, for slabs with in-plane lattice parameters constrained to the bulk value (fix-IP), $d_{111}$ exhibits a constant expansion of $\approx$0.03~\AA~above the bulk value (2.96~\AA). This thickness-independent offset indicates that, under unscreened conditions, $d_{111}$ does not converge to the bulk value with increasing slab thickness. Since this expansion is absent in the nonpolar slab (Fig.~S1) and is independent of thickness, it can be unambiguously attributed to the presence of the depolarization field. This behavior differs fundamentally from conventional perovskite ferroelectrics such as BaTiO$_3$, where imperfect screening leads to a depolarization field that increases inversely with decreasing film thickness due to a nearly thickness-independent interfacial potential drop~\cite{Junquera03p506}. In contrast, HfO$_2$(111) exhibits a hyperferroelectric character, resulting in a nearly thickness-independent depolarization field and $d_{111}$ under different screening conditions (Fig.~S2). When the in-plane lattice is allowed to relax (relax-IP), representing a free-standing film in mechanical equilibrium, $d_{111}$ increases even further. Notably, the magnitude of this additional expansion is inversely correlated with film thickness, qualitatively reproducing the anomalous size effect observed in experiments and confirming its origin in surface stress, an intrinsic mechanical tension arising from the distinct atomic environment at the film surfaces that induces a thickness-dependent in-plane strain beyond the fix-IP constraint (see discussions below). We note that, although surface-related contributions are also present in the fix-IP configuration, their influence is mitigated by excluding the surface layers when evaluating structural quantities such as the out-of-plane lattice spacing $d_{111}$ and the polar displacement $\delta_{111}$.

These findings under OCBC provide several insights. First, they clarify the anomalous size effect: the experimentally observed out-of-plane lattice expansion should not be interpreted as evidence of enhanced polarization or strain-driven ferroelectricity. On the contrary, it coexists with a significantly suppressed polar displacement. Next, the idealized OCBC calculations also enable the separation of the two fundamental mechanisms contributing to this anomalous expansion. One mechanism is a thickness-independent expansion induced by the depolarization field, which is given by $\mathcal{E}_{\rm d} = -P_{\rm s}/\epsilon$, where $P_{\rm s}$ is the spontaneous polarization and $\epsilon$ is the dielectric permittivity. Here, both $\mathcal{E}_{\rm d}$ and $P_{\rm s}$ are determined self-consistently at electrical equilibrium under OCBC and are therefore nearly thickness-independent, analogous to the electric field in an ideal parallel-plate capacitor, which remains constant regardless of plate separation. As discussed below, the depolarization field combined with the negative piezoelectric effect intrinsic to ferroelectric HfO$_2$ contributes to out-of-plane lattice expansion. We note that the depolarization field originates from the incomplete screening of polarization-induced bound charges, reflecting bulk polarization under finite-size boundary conditions rather than a surface-specific chemical effect, as further supported by Fig.~S3, where the polar $Pca2_1$ slab shows a clear potential drop whereas the nonpolar $P4_2/nmc$ slab with identical surface termination exhibits a symmetric potential profile and no depolarization field. The other mechanism is the surface stress effect, which becomes increasingly pronounced as the film thickness decreases; this contribution is explicitly separated from the depolarization-field-driven effect in our analysis. Finally, and most strikingly, a finite polarization persists in free-standing HfO$_2$(111) slabs as thin as 0.6 nm (Fig.~S4). This demonstrates that (111)-oriented HfO$_2$ films fundamentally avoid the conventional critical thickness limit, maintaining robust ferroelectricity even under the most stringent electrostatic boundary conditions~\cite{Sai09p107601}.

\begin{figure}[t]
\includegraphics[width=6.72 in]{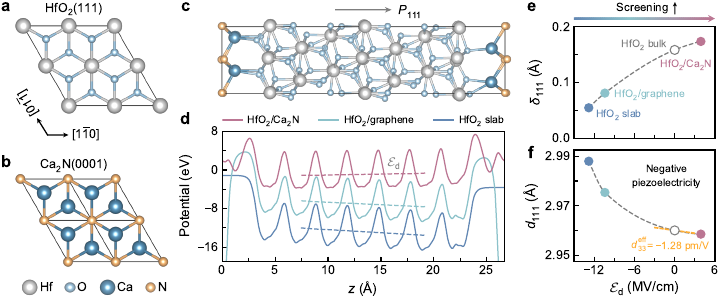}
\caption{\textbf{Screening the depolarization field in HfO$\boldsymbol{_2}$(111) films.} Top view of (\textbf{a}) the high-symmetry cubic HfO$_2$(111) surface and (\textbf{b}) the two-dimensional electride Ca$_2$N. The in-plane lattice of (2$\times$2) Ca$_2$N resembles that of (1$\times$1) HfO$_2$, with Ca and N atoms occupying the O and Hf sites, respectively. (\textbf{c}) Supercell structure of a seven-layer HfO$_2$/Ca$_2$N ferroelectric capacitor viewed along [1$\bar{1}$0], with in-plane lattice mismatches of 1.1\% and $-$0.8\% along [1$\bar{1}$0] and [01$\bar{1}$], respectively. (\textbf{d}) Plane-averaged electrostatic potentials for the HfO$_2$/Ca$_2$N capacitor, HfO$_2$/graphene capacitor, and bare HfO$_2$ slab. The latter two are shifted to align HfO$_2$ layers and improve visibility. Dashed lines indicate macroscopic averages taken over one lattice spacing, with the slope yielding the depolarization field \Ed~(positive direction defined along the polarization). Out-of-plane (\textbf{e}) polar displacement $\delta_{111}$ and (\textbf{f}) lattice spacing $d_{111}$ as functions of \Ed~for different systems. The open circle denotes bulk HfO$_2$ without depolarization. With enhanced screening from the bare slab to the HfO$_2$/Ca$_2$N capacitor, $\delta_{111}$ recovers toward the bulk value, while the reduced $d_{111}$ reflects a negative effective longitudinal piezoelectric response ($d_{33}^{\rm eff}$) along [111].}
\label{fig2}
\end{figure}

To understand the role of the depolarization field, we perform a series of model calculations, which reveal that the $\mathcal{E}_{\rm d}$-induced out-of-plane lattice expansion in HfO$_2$(111) films originates from its unusual negative piezoelectric response. Specifically, we simulate various HfO$_2$(111) capacitor structures with systematically tuned degrees of electrical screening. Through a targeted search of the Computational 2D Materials Database~\cite{Haastrup18p042002}, we identify the two-dimensional electride Ca$_2$N~\cite{Lee13p336} as an ideal electrode, given its compatible hexagonal-like symmetry and excellent lattice match with the HfO$_2$(111) surface (Fig.~\ref{fig2}a-b). We compute the electrostatic potential profiles (Fig.~\ref{fig2}d) for three configurations: a bare HfO$_2$ slab, a slab sandwiched between graphene layers, and a fully assembled HfO$_2$/Ca$_2$N capacitor (Fig.~\ref{fig2}c). In all capacitor models, the in-plane lattice constants are fixed to those of bulk $Pca2_1$ HfO$_2$ projected onto the (111) plane. The resulting lattice mismatch is accommodated as epitaxial strain within the electrode layers, characterized by the in-plane strain components ($\eta_{[1\bar{1}0]}$, $\eta_{ [01\bar{1}]}$) = ($1.1$\%, $-0.8$\%) for the ($2\times2$) Ca$_2$N and ($-1.5$\%, $-3.3$\%) for the ($3\times3$) graphene. These configurations establish a controlled spectrum of depolarization fields, from strong negative fields in the bare slab ($-12.8$ MV/cm), to partial screening with graphene ($-10.4$ MV/cm), and finally to overscreening with Ca$_2$N ($+3.9$ MV/cm), where the screening overcompensates the polarization charge. The superior screening capability of Ca$_2$N is further evident in the layer-resolved density of states (Fig.~S5) and interfacial charge analysis (Fig.~S6), which reveal enhanced charge transfer and a short screening length at the HfO$_2$/Ca$_2$N interface. These interfacial electronic structures may be suggestive of a charge spill-out effect, similar to that discussed by Stengel \etal~for metal/ferroelectric interfaces~\cite{Stengel11p235112}. A detailed investigation of this effect is beyond the scope of the present work, as the HfO$_2$/Ca$_2$N capacitor is employed here primarily as a validation platform providing a controlled range of depolarization fields.

Plotting the structural parameters as a function of $\mathcal{E}_{\rm d}$ reveals the origin of the anomalous lattice expansion (Fig.~\ref{fig2}e-f). As screening improves and \Ed~becomes less negative and approaches zero, the out-of-plane polar displacement, $\delta_{111}$, is progressively restored to its intrinsic bulk value (Fig.~\ref{fig2}e). Simultaneously and counterintuitively, the out-of-plane lattice spacing, $d_{111}$, systematically decreases (Fig.~\ref{fig2}f). This inverse correlation provides direct evidence of a negative effective piezoelectric response along the [111] direction: a stronger polarization leads to a smaller lattice spacing. Such behavior is rooted in HfO$_2$'s intrinsic negative longitudinal piezoelectric coefficient ($d_{33}<0$), which has been previously predicted theoretically~\cite{Liu20p197601} and confirmed experimentally~\cite{Dutta21p7301}. Our own first-principles calculations support this finding, yielding a piezoelectric tensor that, when transformed into the laboratory frame (with $z$ normal to the (111) surface), gives an effective coefficient of $d_{33}^{\rm eff} = -1.02$ pm/V. This value is in excellent agreement with the slope $\partial \eta_{\rm OP} / \partial \mathcal{E}_{\rm d} = -1.28$ pm/V, with out-of-plane strain defined as $\eta_{\rm OP} = (d_{111}-d_{111}^0)/d_{111}^0$, extracted from the linear region of the $d_{111}$–\Ed~relationship near the bulk limit.

We now address the second contribution to the anomalous expansion, which arises from surface stress. As revealed in Fig.~\ref{fig1}c, allowing the in-plane lattice of the OCBC slab to relax induces an additional out-of-plane expansion, the magnitude of which is inversely correlated with film thickness. This thickness-dependent behavior can be quantitatively explained by a surface elasticity model~\cite{Lv21p197403}. The total elastic energy, $U_{\rm slab}$, of a free-standing film is the sum of its bulk strain energy and the energy of its two surfaces:
\begin{equation}
U_{\rm slab} = U_{\rm bulk} + 2A\gamma = \dfrac{1}{2}C\eta_{\rm IP}^2At + 2A(\sigma\eta_{\rm IP}+\dfrac{1}{2}S\eta_{\rm IP}^2),
\label{eq1}
\end{equation}
where the first term represents the bulk energy under an in-plane biaxial strain $\eta_{\rm IP}$ (with $A$ being the area, $t$ the thickness, and $C$ the effective elastic constant), and the second term describes the surface energy, with $\sigma$ and $S$ being the effective surface stress and surface elastic constant, respectively~\cite{Schmid95p10937}. Crucially, the parameters $\sigma=0.38$ eV/${\rm \AA}^2$ and $S=-4.78$ eV/${\rm \AA}^2$ are fitted directly from our OCBC slab calculations at a thickness of four layers (Fig.~S7). Therefore, these are effective parameters that implicitly capture the influence of the depolarization field on the surface energetics, including both structural and electronic responses of the film under OCBC, most notably the suppression of polarization.

\begin{figure}[t]
\includegraphics[width=4.76 in]{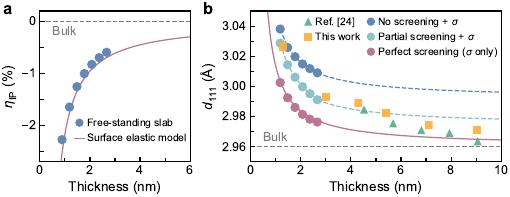}
\caption{\textbf{Effects of surface stress and depolarization-field screening on out-of-plane lattice spacing.} (\textbf{a}) Surface-stress-induced in-plane strain ($\eta_{\rm IP}$) as a function of film thickness ($t$). The surface elastic model gives $\eta_{\rm IP}(t)=\dfrac{-2\sigma}{Ct+2S}$, with bulk elastic constant $C=4.44$ eV/${\rm \AA}^3$, surface stress $\sigma=0.38$ eV/${\rm \AA}^2$, and surface elastic constant $S=-4.78$ eV/${\rm \AA}^2$. A positive $\sigma$ yields a compressive in-plane strain. The model predictions (purple line) match well with DFT-calculated values (blue circles). (\textbf{b}) Thickness-dependent out-of-plane lattice spacing $d_{111}$ under perfect, partial, and no screening of the depolarization field. Perfect, partial, and no screening are represented by bulk HfO$_2$, HfO$_2$/graphene capacitor, and free-standing HfO$_2$ slab, respectively. In the hypothetical case of perfect screening, the lattice expands due to the Poisson effect, driven by in-plane strain induced by surface stress ($\sigma$). The dashed lines for the no-screening and partial-screening cases at larger thickness are schematic guides indicating the asymptotic regime, where the depolarization-induced lattice expansion approaches an approximately thickness-independent offset relative to the perfectly screened limit. Experimental $d_{111}$ values of ferroelectric Hf$_{0.5}$Zr$_{0.5}$O$_2$ films grown on SrTiO$_3$(001) substrates are taken from Ref.~\cite{Lyu25p32596} as well as from the present work.}
\label{fig3}
\end{figure}

For a free-standing film, the system relaxes to an equilibrium strain that minimizes the total elastic energy in Eq.~\ref{eq2}. Setting the derivative $\partial U_{\rm slab} / \partial \eta_{\rm IP}$ to zero yields the surface-stress-induced in-plane strain (relative to the bulk lattice constant) as a function of thickness:
\begin{equation}
\eta_{\rm IP}(t) = \dfrac{-2\sigma}{Ct+2S}.
\label{eq2}
\end{equation}
Since the DFT-calculated surface stress $\sigma$ is positive, the model predicts a compressive in-plane strain that is most pronounced in ultrathin films and vanishes with increasing thickness. As shown in Fig.~\ref{fig3}a, the analytical expression closely matches the DFT-calculated in-plane strains of free-standing slabs across varying thicknesses, confirming the role of surface stress. This, in turn, contributes to the thickness-dependent component of the out-of-plane lattice expansion. We note that, in realistic thin films, the formation of different interfaces can introduce different interface stress effects; therefore, the surface stress considered in these free-standing slab configurations represents an idealized limit in which surface-stress contributions are fully active.

Based on the preceding analysis, we develop a general model that quantitatively describes the anomalous out-of-plane lattice expansion under varying degrees of electrical screening. The total out-of-plane strain, $\eta_{\rm OP}$, for a film of thickness $t$ can be decomposed into two distinct contributions:
\begin{equation}
\eta_{\rm OP}(t) = -\nu \eta_{\rm IP}(t) + \eta_{\rm OP}(\mathcal{E}_{\rm d}).
\label{eq3}
\end{equation}
The first term, $-\nu \eta_{\rm IP}(t)$, represents the elastic response via the Poisson effect, where $\nu$ is the Poisson's ratio and $\eta_{\rm IP}(t)$ is the in-plane strain induced by surface stress (as expressed in Eq.~\ref{eq2}). This term is inherently thickness-dependent. The second term, $\eta_{\rm OP}(\mathcal{E}_{\rm d})$, accounts for the out-of-plane expansion from the negative piezoelectric effect, which depends on the strength of the residual depolarization field, $\mathcal{E}_{\rm d}$, and is thus determined by the screening environment.

This framework allows us to dissect the contributions, as shown in Fig.~\ref{fig3}b. The no-screening case (blue line), corresponding to our fully relaxed free-standing OCBC slabs, represents the upper bound, where both surface stress and the maximum negative piezoelectric effect contribute. We can isolate the surface-stress effect by modeling a perfect screening scenario ($\mathcal{E}_{\rm d} = 0$), where the piezoelectric term vanishes. For a given thickness, this is computationally realized by applying the in-plane strain $\eta_{\rm IP}(t)$, extracted from a free-standing slab of the same thickness, to a bulk HfO$_2$ crystal (free of depolarization field), and relaxing it along the out-of-plane direction. The resulting expansion (purple line) reflects purely the Poisson effect and establishes the lower bound for the lattice spacing. We then model an intermediate case of partial screening by imposing the same in-plane strain to a HfO$_2$/graphene capacitor, which sustains a reduced but non-zero \Ed. The resulting curve (cyan line) lies neatly between the perfect- and no-screening limits. Remarkably, the experimental data in Ref.~\cite{Lyu25p32596} and our measurements indicate that the observed thickness dependence of $d_{111}$ falls within the range bounded by our theoretical predictions for the free-standing slab (no screening) and the bulk (perfect screening) limit. This provides strong evidence that the observed lattice expansion in experimentally fabricated ferroelectric hafnia films arises from a combined effect of thickness-dependent surface stress and a negative piezoelectric response to an imperfectly screened depolarization field.

Finally, the remarkable stability of ferroelectricity in HfO$_2$(111) films under OCBC suggests that HfO$_2$ exhibits orientation-dependent hyperferroelectricity. A hyperferroelectric is defined by its ability to maintain a stable, spontaneous polarization even under OCBC~\cite{Garrity14p127601}. The LGD free energy model for a ferroelectric film under OCBC is given as (see the Supplemental Material for the derivation)~\cite{Garrity14p127601,Li16p34085,Adhikari19p104101}:
\begin{equation}
F(\lambda) = U(\lambda) - \Omega(\lambda) \left[ \dfrac{1}{2}P_{\rm s}(\lambda)\mathcal{E}_{\rm d}(\lambda) + \dfrac{1}{2}\epsilon_0\chi_{\rm e}(\lambda)\mathcal{E}_{\rm d}^2(\lambda) \right],
\label{eq4}
\end{equation}
where $U(\lambda)$, $\Omega(\lambda)$, $P_{\rm s}(\lambda)$, $\mathcal{E}_{\rm d}(\lambda)$, and $\chi_{\rm e}(\lambda)$ are, respectively, the total energy per unit cell, the unit-cell volume, the spontaneous polarization, the depolarization field, and the electronic susceptibility at the configuration of ferroelectric distortion $\lambda$; $\epsilon_0$ is the electric permittivity of free space. Imposing the vanishing displacement field under OCBC, $D = \epsilon_0 \left[ 1+\chi_{\rm e}(\lambda) \right]\mathcal{E}_{\rm d}(\lambda) + P_{\rm s}(\lambda) = 0$, yields the depolarization field $\mathcal{E}_{\rm d}(\lambda) = -P_{\rm s}(\lambda) / \epsilon_0\left[1+\chi_{\rm e}(\lambda)\right]$. Substituting this back into the free energy expression gives:
\begin{equation}
F(\lambda) = U(\lambda) + \dfrac{\Omega(\lambda)P_{\rm s}^2(\lambda)}{2\epsilon_0\left[1+\chi_{\rm e}(\lambda)\right]^2},
\label{eq5}
\end{equation}
where the second term is the depolarization energy penalty.

\begin{figure}[t]
\includegraphics[width=4.76 in]{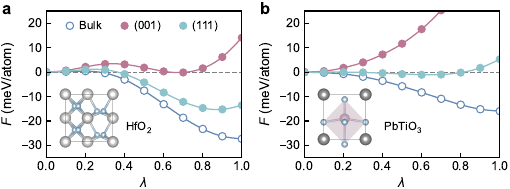}
\caption{\textbf{Orientation-induced hyperferroelectricity in HfO$\boldsymbol{_2}$ and PbTiO$\boldsymbol{_3}$ films.} Free energy ($F$) of thin films as a function of the ferroelectric distortion parameter ($\lambda$) in (\textbf{a}) HfO$_2$ and (\textbf{b}) PbTiO$_3$, taking the energy of the paraelectric $P4_2/nmc$ and $Pm\bar{3}m$ phases as reference, respectively. Filled markers denote the free energy under open-circuit boundary conditions, as given in Eq.~\ref{eq5}. Insets show the unit cells of ferroelectric HfO$_2$ and PbTiO$_3$.}
\label{fig4}
\end{figure}

To construct the free-energy profiles, $U(\lambda)$, $\Omega(\lambda)$, $P_{\rm s}(\lambda)$, and $\chi_{\rm e}(\lambda)$ are obtained from DFT calculations along a continuous ferroelectric distortion path connecting the paraelectric $P4_2/nmc$ phase and the ferroelectric $Pca2_1$ phase. These quantities are then directly used in Eq.~\ref{eq5} to evaluate the orientation-dependent free energy landscapes. Figure~\ref{fig4}a shows the free energy of HfO$_2$ as a function of ferroelectric distortion $\lambda$. For bulk HfO$_2$, the free energy reduces to $F(\lambda)=U(\lambda)$, since no depolarization contribution is present. For HfO$_2$(001) films, the polarization is purely out-of-plane, maximizing the depolarization energy penalty. This penalty completely overwhelms the intrinsic double-well potential with a well depth of 27 meV/atom, stabilizing the nonpolar phase as the ground state. This agrees with previous theoretical studies, which found that relaxed HfO$_2$(001) stoichiometric slabs adopt either a nonpolar $P2_1/c$-like structure~\cite{Acosta23p124401} or an antipolar $Pbcn$ phase~\cite{Lv25p084407}. In contrast, for the (111) orientation, only the [111] component of $P_{\rm s}$ contributes to the depolarization energy. The resulting free energy of HfO$_2$(111) films under OCBC exhibits a shallower, yet still distinct, double-well potential that preserves the ferroelectric ground state.

To test the generality of this principle, we apply the same analysis to the prototypical soft-mode ferroelectric, PbTiO$_3$ (Fig.~\ref{fig4}b). As expected, the model predicts that a PbTiO$_3$(001) film becomes nonpolar under OCBC. However, for a PbTiO$_3$(111) film, the reduced depolarization penalty is predicted to allow for a stable polarized ground state. We confirmed this prediction by performing direct DFT calculations on a PbO$_3$-terminated PbTiO$_3$(111) slab of 0.9 nm, which indeed relaxes to a structure with robust out-of-plane polar displacements (Fig.~S8). These results strongly suggest that orientation engineering is a viable strategy for realizing hyperferroelectricity. More generally, previous studies have demonstrated that surface orientation can strongly modify ferroelectric and piezoelectric responses~\cite{Raeliarijaona14p054105,Yang24p95}. By selecting a crystallographic orientation that reduces out-of-plane polarization, ferroelectric thin films can retain polarization without screening electrodes, effectively eliminating the critical thickness limit.

In summary, this work resolves the puzzle of the anomalous size effects in ferroelectric HfO$_2$ thin films. The experimentally observed out-of-plane lattice expansion is not a sign of enhanced ferroelectricity, but instead coexists with suppressed polar displacements. This expansion originates from the interplay between two distinct mechanisms: a negative longitudinal piezoelectric response to the residual depolarization field, and a positive surface stress that becomes increasingly pronounced at reduced thickness. The (111)-orientation is found to enable HfO$_2$ to sustain a stable polarization even under open-circuit boundary conditions, revealing a form of orientation-dependent hyperferroelectricity. This concept generalizes beyond HfO$_2$, offering a pathway to eliminate the critical thickness limit in conventional perovskite ferroelectrics and enabling robust ferroelectricity in ultrathin films. Moreover, the identification of the two-dimensional electride Ca$_2$N as a nearly ideal electrode material, capable of effectively screening the depolarization field and restoring bulk-like ferroelectric properties in nanometer-scale capacitors, provides a concrete materials solution for hafnia-based ferroelectric device applications.

\section*{Methods}

\subsection*{Density functional theory calculations}
Density functional theory (DFT) calculations were performed using the Vienna \textit{ab initio} simulation package (VASP)~\cite{Kresse96p11169} with the projector augmented-wave (PAW) method~\cite{Blochl94p17953,Kresse99p1758}. The Perdew--Burke--Ernzerhof (PBE)~\cite{Perdew96p3865} and PBE for solids (PBEsol)~\cite{Perdew08p136406} exchange correlation functionals were employed for HfO$_2$ and PbTiO$_3$, respectively. The plane-wave cutoff was set to 600 eV. Brillouin zones were sampled with $\Gamma$-centered Monkhorst--Pack $k$-point meshes~\cite{Monkhorst76p5188}: $4\times4\times4$ for HfO$_2$, $6\times6\times6$ for PbTiO$_3$, and $3\times3\times1$ for HfO$_2$(111) slabs and capacitors. All structures were optimized until the residual atomic forces were below 0.01 eV/\AA. Dipole corrections~\cite{Neugebauer92p16067} were applied along the surface normal in slab calculations. Polarization was determined using the Berry phase approach~\cite{KingSmith93p1651,Vanderbilt93p4442}, and piezoelectric coefficients were obtained via density functional perturbation theory (DFPT)~\cite{Wu05p035105}. Atomic structures were visualized with VESTA~\cite{Momma11p1272}.

\subsection*{Pulsed laser deposition of HZO thin films}
Epitaxial Hf$_{0.5}$Zr$_{0.5}$O$_2$ (HZO) thin films of various thicknesses were grown on La$_{0.67}$Sr$_{0.33}$MnO$_3$ (LSMO)-buffered SrTiO$_3$ (001) single-crystal substrates by pulsed laser deposition (Arrayed Materials RP-B system). A 10-nm-thick LSMO bottom electrode was first deposited at 700~\textcelsius~with a laser repetition rate of 3~Hz and a fluence of 0.8~J/cm$^2$. The ferroelectric HZO layer was subsequently deposited at 600~\textcelsius~using a repetition rate of 2~Hz and a fluence of 1.3~J/cm$^2$. After deposition, the samples were cooled to room temperature at a rate of 10~\textcelsius~per minute under a static oxygen pressure of 10$^4$~Pa. For electrical characterization, circular platinum (Pt) top electrodes (12.5~$\mu$m in diameter, $\sim$100~nm in thickness) were patterned by photolithography and deposited via magnetron sputtering (Arrayed Materials RS-M system).

\section*{Acknowledgements}
T.Z. and S.L. acknowledge the supports from National Key R\&D Program of China (Grant No. 2021YFA1202100), National Natural Science Foundation of China (Grant No. 12361141821, 12074319, 12404114), and Zhejiang Provincial Natural Science Foundation of China (LR25A040004). The computational resource is provided by Westlake HPC Center.

\clearpage

\clearpage

\clearpage

\clearpage

\clearpage
\bibliography{SL}

\end{document}